\documentclass[twocolumn,aps,prd,   
               preprintnumbers,numbers,sort&compress,
               nofootinbib,
                            showpacs,
               colorlinks,
               linkcolor=blue,
               citecolor=blue]{revtex4-1} 

\usepackage{graphicx}
\usepackage{dcolumn}
\usepackage{bm}
\usepackage{hyperref}

\usepackage{graphicx}
\usepackage{dcolumn}
\usepackage{bm}
\usepackage{hyperref}

\usepackage{amsmath}
\usepackage[caption=false]{subfig}

 \def\ra{\rangle}
\def\la{\langle}

\newcommand{\be}{\begin{eqnarray}}
\newcommand{\ee}{\end{eqnarray}}
\newcommand{\beq}{\begin{equation}}
\newcommand{\eeq}{\end{equation}}
\newcommand{\exclude}[1]{}
\newcommand{\Red}[1]{{\color{red}({#1})}}

\exclude{

}
\graphicspath{{./Figures/}} 

\begin{document}

\title{Telescope Array Bursts, Radio Pulses and Axion Quark Nuggets}


\author{Xunyu Liang 
     and        Ariel  Zhitnitsky }
 \affiliation{Department of Physics and Astronomy, University of British Columbia, Vancouver, V6T 1Z1, BC, Canada}

\begin{abstract}
Telescope Array (TA) experiment has recorded  \cite{Abbasi:2017rvx,Okuda_2019}  several short time bursts of air shower like events. 
These bursts are very distinct from conventional single showers, and  are found to be strongly correlated with lightnings. 
In our previous work \cite{Zhitnitsky:2020shd} we proposed that these bursts  represent  the direct manifestation of the
dark matter (DM) annihilation events within  the so-called axion quark nugget (AQN) 
model. In the present work we suggest   to test this proposal to search for  the    radio signals in  frequency band $\nu\in (0.5-200)$  MHz   which must be synchronized with the TA bursts.  We argued that the conventional lightning-induced radio emission can be easily discriminated from the AQN-induced radio pulses discussed in this work. 
\end{abstract}

\maketitle

\section{Introduction}
\label{sec:introduction}

This work is tightly linked to our recent proposal \cite{Zhitnitsky:2020shd} interpreting the mysterious bursts  observed  by Telescope Array (TA) experiment  \cite{Abbasi:2017rvx,Okuda_2019} in terms of the AQN annihilation events under the thunderstorm.  These events are very unusual and cannot be interpreted  in 
terms of  conventional single showers as reviewed below.  In the present work we shall argue that the proposed mechanism \cite{Zhitnitsky:2020shd} in terms of  the AQN annihilation events inevitably predict the radio wave pulses which must be synchronized with TA bursts.  Based on this prediction we suggest   to test  our  proposal by searching  for  the     radio signals in  frequency band $\nu\in (0.5-200)$ MHz  which must be synchronized with the TA bursts, which represents the main goal of the present work.   

Such test would unambiguously support or refute the proposal (interpreting TA bursts as the AQN events) because the main arguments of the  present work based  on  synchronization which represents pure geometrical property  of the system. This is  because the radio pulse and the TA burst are originated from the same location, emitted at the same instant, and propagate with the speed of light.  
This feature  of synchronization is not sensitive to many assumptions and uncertainties 
which were inevitably  present  in estimates  \cite{Zhitnitsky:2020shd} interpreting the TA bursts as the AQN annihilation events. Furthermore, we also argue that the conventional lightning-induced radio emission is qualitatively different from the AQN -induced radio pulses discussed in this work. 
Therefore,  these two different radio signals can be easily distinguished from each other, and we advocate to conduct such studies.

Finally, before we proceed one should emphasize from the start that AQN events are generically random events as   they are related to DM particles. However, as mentioned in \cite{Zhitnitsky:2020shd}  the AQN hitting the thunderstorm area may serve as a trigger which sparks the lightning. Therefore, the AQNs play the dual role in our proposal. First,  they    play the same role  of  the CRs which are normally assumed to initiate the lightning processes. Secondly, they emit highly energetic particles being recorded as TASD bursts. 
This represents a key element why mysterious bursts \cite{Abbasi:2017rvx,Okuda_2019} which are associated with lightning events can be indeed originated from the AQNs  which happen to appear in the area under the thunderclouds. The event rate estimated in \cite{Zhitnitsky:2020shd} is based on this assumption.

Our presentation is organized as follows. In next Sect. \ref{AQN} we overview the basic ideas of the AQN model, in Sect. \ref{TA-overview} we overview the TA bursts  observations \cite{Abbasi:2017rvx,Okuda_2019} and the basic ideas  of the proposal \cite{Zhitnitsky:2020shd} with  the   resolution of these puzzling bursts  in terms of the AQN annihilation events. 
In Sect.     \ref{radio} and \ref{numerics}  we argue that the emission of the radio pulse is inevitable consequence of the proposed  mechanism, and we estimate its numerical characteristics. 
In Sect. \ref{sec:Geosynchrotron radiation} and \ref{B-numerics} we estimate the radio pulse due to the Earth's magnetic field.   
In Sect. \ref{mimic} we explain how  the  conventional radio signals induced by  the lightnings   could be easily 
discriminated from the AQN-induced radio pulses. 
Finally, Sect. \ref{conclusion} is our conclusion where we  suggest to  search for a synchronized radio pulse  with TA burst.

\section{The AQN DM model}
\label{AQN}
The AQN model was invented two decades ago \cite{Zhitnitsky:2002qa}  
 to   explain  the observed  similarity between the dark matter and the visible matter  densities in the Universe\footnote{Note that various names have been used since the early development of the model, such that AQN might be referred as the QCD ball, quark nugget, compact composite objects (CCO), etc. in the past literature.}. In this model, DM is made out of macroscopic lumps of quarks (or antiquarks) in colour superconducting (CS) phase with characteristic mass and size of order grams and $0.1\rm{\,\mu m}$ respectively. In the AQN  framework the 
baryogenesis is actually a charge segregation  (rather than charge generation) process 
when  the global baryon number of the universe remains 
zero at all times.   Formation of AQNs relies on the coherent $\cal CP$-violating axion field in early Universe, and consequently it leads to an asymmetry between matters and antimatter nuggets, which results in asymmetry between visible matter and  antimatter  in the observable Universe.   

Unlike many conventional DM candidates such as weakly interacting massive particles (WIMPs) and axion, the AQN model naturally explains the observed similarity between the dark and visible densities in the Universe, i.e. $\Omega_{\rm DM}\sim\Omega_{\rm vis}$, with no fitting parameters, as both DMs and visible matters share the same QCD origin. As its name suggests, the AQN is similar to Witten's quark nugget, see \cite{Witten:1984rs,Farhi:1984qu,DeRujula:1984axn},  and    review \cite{Madsen:1998uh}, in many respects. This type of DM is in fact strongly-interacting by virtue of its macroscopic size, but  it serves as the DM  due to its diminutive number density.

In contrast to Witten's original proposal, formation of AQNs does not rely on a first order QCD phase transition. Rather, accumulation and compression of baryon (or anti-baryon) charge come from oscillations   of axion domain wall (DW) bubbles during the QCD transition in early Universe which can squeeze the quark matter inside the bubbles. Due to the   surface tension of the DW, ultimately the baryon (or antibaryon) charge trapped in a bubble is squeezed into a quark (or antiquark) nugget in CS phase. Because the CS phase is far more stable than the conventional hadronic phase, AQNs   do not suffer from the evaporation problem, in contrast with the original  Witten's proposal \cite{Witten:1984rs,Farhi:1984qu,DeRujula:1984axn,Madsen:1998uh}. We refer to the original papers \cite{Liang:2016tqc,Ge:2017ttc,Ge:2017idw,Ge:2019voa} and a brief review \cite{Zhitnitsky:2021iwg} devoted to the specific questions related to the AQN's formation, generation of the baryon asymmetry, and survival pattern in early Universe. We also refer  to 
  independent analysis \cite{Santillan:2020lbj} supporting the basic elements on the formation and survival pattern of the AQNs during the early stages of the evolution,  including the BBN and CMB epochs.    
  
  For the present studies, however,  we take the  agnostic viewpoint, and assume that such nuggets made of antimatter are present in our Universe today irrespective to their   formation mechanism. This assumption is consistent with all presently available cosmological, astrophysical and terrestrial  constraints as long as  the average baryon charge of the nuggets is sufficiently large as we review  below.

The AQN DM model is consistent with all presently available cosmological, astrophysical and terrestrial constraints. The strongest direct detection limit is set by the IcuCube observatory, see Appendix A in \cite{Lawson:2019cvy}:
\begin{equation}
\label{direct}
\la B \ra > 3\times 10^{24}\qquad({\rm direct ~ non-detection ~constraint)}\,.
\end{equation}
Similar limits are from the ANITA experiment and geothermal constrains which are also consistent with (\ref{direct}) as estimated in \cite{Gorham:2012hy}. One soft constraint comes from the indirect non-detection of etching tracks in ancient mica \cite{Jacobs:2014yca}, as it gives a more stringent  limit in the available parameter space of DM nuggets with mass $M>55{\rm\,g}$, or correspondingly $B\gtrsim10^{25}$ in the AQN model.  However this constraint is based on an oversimplified assumption that all nuggets have equal mass, which is invalid in the AQN model since the AQN's formation mechanism implies a reasonably broad distribution of nugget's size. It is also claimed in \cite{Gorham:2015rfa} that the AQNs cannot account for more than 20$\%$ of the DM density according to the neutrino flux limits in the 20-50 MeV range observed from Super-Kamiokande (SuperK). However, as pointed  out in \cite{Lawson:2015cla}, the claim \cite{Gorham:2015rfa} is based on an incorrect assumption that AQNs produce a neutrino spectrum similar to the conventional baryon-antibaryon annihilation events\footnote{In conventional baryon-antibaryon annhilation events, pions and muons are cospiously produced and consequenly generate a signigicant number of neutrinos and antineutrinos with energy 20-50 MeV that meets the high sensitivity range of SuperK.}. Rather, the annihilation processes in CS phase are dramatically different from the conventional scenarios \cite{Lawson:2015cla} because the energy scale of  the lightest pseudo Goldstone mesons (pions and kaons) is in 20 MeV range, rather than 140 MeV in hadronic phase. The  resulting neutrino spectrum features  in CS phase computed in \cite{Lawson:2015cla,Zhitnitsky:2019tbh} are consistent with the observations.

The presence of the {\it antimatter} nuggets in the system implies that there will be  annihilation events leading to  large number of observable effects on different scales: from galactic scales to the terrestrial rare events. In fact, there are many hints suggesting that such annihilation events may  indeed  took place in early Universe as well as they are happening now in present epoch. In particular, the  AQNs might be responsible for a resolution of   the  ``Primordial Lithium Puzzle" \cite{Flambaum:2018ohm}
 during BBN epoch. The AQNs  may also  alleviate the tension between standard model cosmology and the recent EDGES observation of a stronger than anticipated 21 cm absorption feature as argued in \cite{Lawson:2018qkc}.  
 The AQNs may also explain  
  a recently observed   ``exotic" diffuse UV radiation in our galaxy 
 \cite{Henry_2014} with very puzzling features, which are  very hard to interpret within conventional astrophysical models \cite{Zhitnitsky:2021wjb}. 
 The AQNs might be also responsible for famed  long standing problem of  the ``Solar Corona Mystery"
  \cite{Zhitnitsky:2017rop,Raza:2018gpb} when the   so-called ``nanoflares" conjectured by Parker long ago \cite{Parker} are   identified with the  annihilation events in the AQN framework. The AQNs could be also responsible for other mysterious and anomalous CR like events (along with \cite{Zhitnitsky:2020shd} we already mentioned). It includes  a  mysterious   anomalous events with noninverted polarity  observed by the  Antarctic Impulse Transient Antenna (\textsc{ANITA})    collaboration \cite{Liang:2021rnv},   and  Multi Modal Clustering anomalous events  \cite{Zhitnitsky:2021qhj} observed by the HORIZON 10T collaboration.

We conclude this short overview of the AQN model by emphasizing that the model is consistent with all present  constraints   as long as average baryon charge of the nuggets satisfies the relation (\ref{direct}).  Furthermore, with the same set of parameters within the same framework it  may explain a large number of mysterious phenomena listed above, which apparently suggest that the DM might be indeed in form of the mater and antimatter quark nuggets. 
In what follows we  assume that such antimatter nuggets exist and together with nuggets saturate the DM density  today.

\section{TA bursts as the AQN annihilation events under thunderclouds}
\label{TA-overview}

The TA is designed for detection of extensive air showers induced by ultrahigh energy cosmic rays (CRs) consisting of 507 ground surface particle detectors (SDs) and 3 atmospheric fluorescence telescope stations. The TA collaboration have reported 10 excessively unlikely bursts of CR-like events are observed by the SDs \cite{Abbasi:2017rvx,Okuda_2019}. Comparing to conventional CRs, the TA burst events:
\begin{enumerate}  
	\item  have a much smaller dispersion in reconstructed air shower fronts [see Fig. 3 and Fig. 4 in \cite{Abbasi:2017rvx}), and do not have sharp edges in waveforms, the so-called {\it ``curvature" puzzle};
	\item are temporally clustered within 1 ms that is highly unlikely (chance coincidence less than $10^{-4}$ for five-year observation) for ultrahigh energy CRs in the fitted energy range $(10^{18}-10^{19})\,$eV, the so-called {\it ``clustering" puzzle};
	\item are all recorded under thunderstorm, and most of them are ``synchronized'' (less than 1 ms) or ``related'' (less than 200 ms) with the lightning events,
	the so-called {\it ``synchronization" puzzle},
\end{enumerate}
see original articles \cite{Abbasi:2017rvx,Okuda_2019} and short reviews in \cite{Zhitnitsky:2020shd,Zhitnitsky:2021iwg} for technical details.

Proposal \cite{Zhitnitsky:2020shd} suggested the TA bursts can be a natural consequence of an \textit{antimatter} AQN traversing through the thundercloud, where about $10^9$ weakly bound positrons are emitted from the AQN instantaneously in presence of the strong intracloud electric field ($\sim\,$kV/cm) at altitude above 10 km. 

The key mechanism in proposal \cite{Zhitnitsky:2020shd} may be summarized as follows.  First, the electric field $\cal{E}$ in the thunderclouds is characterized by the following parameters \cite{Gurevich_2001,DWYER2014147}
\be
\label{parameters1}
{\cal{E}}\simeq  {\rm \frac{kV}{cm}}, ~~~~  l_a \simeq 100 ~{\rm m},~~~~ \tau_{\cal{E}}\simeq \frac{l_a}{c}\simeq 0.3 {\rm\,\mu s}
\ee  
where $l_a$ is the so-called avalanche length. The electric field is sufficiently strong to ionize an \textit{antimatter} AQN and induce liberation of weakly bound positrons:
\be
\label{Delta_E}
\Delta E \simeq [e{\cal{E}}\cdot R_{\rm cap}]\sim 2 \rm ~keV\gtrsim E_{\rm bound}\,,
\ee
where $E_{\rm bound}\sim\rm keV$ is the binding energy of the positrons, and $R_{\rm cap}\sim2{\rm\,cm}$ is the typical distance (from the nugget's core where positrons reside \cite{Zhitnitsky:2020shd}. 

These liberated positrons will be accelerated to MeV energies in the background of electric field characterized by typical length scale $l_a \simeq 100$ m according to (\ref{parameters1}):
\be
\label{E_MeV}
E_{\rm exit}\simeq [e{\cal{E}} \cdot l_a]\sim 10 \rm ~MeV\,.
\ee
Thereafter, the positrons exit the region of strongly fluctuating electric field which is known to be present under thunderclouds. 

\begin{figure}[ht]
	\centering
	\includegraphics[width=0.9\linewidth]{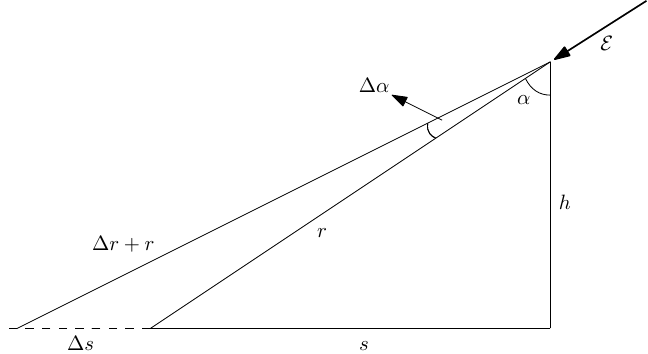}
	\caption{The positrons move along the cone  with angle $\Delta \alpha$ and  inclination angle $\alpha$ with respect to the vertical direction.  The angular spread  $\Delta \alpha\ll \alpha$ is assumed to be small. The spatial spread on the surface is determined  by $\Delta s$, while the additional travelling path is determined by $\Delta r$, see estimates in the text. The altitude is assumed to be within conventional range $ h\simeq (4-12)$ km. Instant  direction of the electric field $\mathbf{\cal{E}}$ at the moment of exit of the positrons   is also shown.  Figure adapted from \cite{Zhitnitsky:2020shd}. }
	\label{geometry}
\end{figure}

Comparison to conventional CR air shower, the positron flux induced by AQN has a smaller dispersion angle $\Delta\alpha$ and spatial spread $\Delta s$: \cite{Zhitnitsky:2020shd}
\begin{equation}
\label{spread2}
\begin{aligned}
&\Delta s \simeq  r \left(\frac{\Delta \alpha}{\cos \alpha}\right) \simeq \frac{1~{\rm km}}{\cos\alpha} \left(\frac{r}{10 \rm ~km}\right)   \left(\frac{\Delta \alpha}{0.1}\right)\,,  \\
&\Delta r \simeq \Delta s \sin\alpha , ~~~\Delta \alpha\simeq \left(\frac{v_{\perp}}{c}\right)\in (0- 0.1)\,,
\end{aligned}
\end{equation}
where we choose the transverse component (with respect to the electric field) of the velocity $v_{\perp}\simeq \sqrt{2 \Delta E/m}\simeq 0.1c$, see Fig. \ref{geometry} for precise definitions of the parameters. 
This behaviour\footnote{One should comment here that the actual trajectories of the positrons are not the straight  lines due to the Earth's magnetic field ${\cal{B}}\sim 0.5$ gauss as we discuss in Sect. \ref{sec:Geosynchrotron radiation}. However, the main point here is that   the  emitted  positrons are characterized by the same velocity and   localization at the instant of exit. Therefore   the presence of the magnetic field ${\cal{B}}$  does not spoil our estimates for the temporal and spatial spreads given  by   Eqs. (\ref{spread2}) which must hold irrespective of the presence of ${\cal{B}}$.} is  consistent with feature {\bf 1} coined as the {\it ``curvature" puzzle}. 

 A TA burst in the AQN framework represents the  cluster of events related to one and the same AQN which traverses through the distance $L_{\rm burst}\simeq250{\rm\,m}$ during $\Delta t\simeq1{\rm\,ms}$ at relativistic speed, rather than a ``coincidental'' combination of independent CR events. It also resolves the dramatic inconsistency (when interpreted in terms of the CR air showers) for the burst events when the event rate favours an energy range in $10^{13}\,$eV, but the intensity of the events suggests $(10^{18}-10^{19})\,$eV, which explains the item {\bf 2} coined as the {\it ``clustering" puzzle.}

Lastly, an AQN can serve  as a trigger of runaway breakdown avalanche as discussed in \cite{Zhitnitsky:2020shd} \footnote{\label{trigger}It is known that the presence of a strong electric field (\ref{parameters1}) is required to initiate a runaway breakdown avalanche. Another element which	remains to be a matter of debates \cite{Gurevich_2001,DWYER2014147,Dwyer:2011lep,Gurevich:2012col,Dwyer:2012rtc,Gurevich:2013rba,Hare:2017dcr} is the nature of the seeded particles which play the role of a trigger to an avalanche. The proposal \cite{Zhitnitsky:2020shd} suggests that AQNs may also play the role of a trigger 
	along with conventional sources such as the CR   \cite{Gurevich_2001,Gurevich:2013rba}.}.
	This explains  the strong correlation between bursts and lightning events   within AQN interpretation. The AQN-induced burst may or may not  initiate lightnings depending on presence of  other required ingredients for runaway breakdown avalanche to start. 
	
   	To rephrase it, an AQN plays the dual role when it propagates in  the thunderclouds:  it emits the very energetic  positrons, and it may also    trigger the lightning, similar to conventional CR. The energetic positrons is the source for   TASD bursts, while the feature of triggering   explains the observed correlation between the bursts and lightnings in the AQN framework.   It explains feature {\bf 3} coined as the {\it ``synchronization" puzzle}, as well as it answers the question of why most but not all bursts are correlated with lightnings.  

\exclude{
The emitted positrons produce a CR-like event, but has the following unique features: 
\begin{enumerate}[label=\textbf{\arabic*}.]
	\item highly concentrates along the central axis and much smaller spatial spread than ultrahigh energy CRs, non-sharp edges in waveforms; 
	\item the burst is not a combination of independent events, {\bf  but in fact represents the cluster being} induced by one and the same AQN (typical time scale 1 ms);
	\item has to be strongly correlated to thunderstorm, but not necessarily correlated to the lightning events, according to the proposed mechanism.
\end{enumerate}

Hence, the features of the AQN-induced bursts fully agree with the ones observed by TA even based on the simplest modeling and estimate.
}
To summarize this section:  the  mysterious  bursts (with highly unusual features {\bf 1-3} as listed at the beginning of the section)  CR-like events observed by TASD  \cite{Abbasi:2017rvx,Okuda_2019}  are naturally interpreted as the cluster events generated by the AQNs propagating in thunderstorm environment. 
Some features of these events (such as intensity and the basic normalization factors) suffer   inevitable  uncertainties due to  very complex dynamical properties of the  system (the AQN propagation under thunderstorm with large Mach number).   However, there are some features of the system such as given by Eqs. (\ref{spread2}) which are not sensitive to these uncertainties and represent almost model-independent consequences of the proposal   \cite{Zhitnitsky:2020shd}.

\section{Radio pulse and TA burst as synchronized events}\label{radio} 
This section is devoted to studies of the radio signals which always accompany TA bursts when interpreted in terms of the AQN annihilation events under the thunderstorm as presented in previous Sect. \ref{TA-overview}. We shall argue below that the emission of the radio pulse is inevitable consequence of the proposed  mechanism. Furthermore, the radio pulse must be  synchronized with TA burst irrespective of   whether the bursts are  related or unrelated    to the lightning events. This  synchronization must be within $10 {\rm\,\mu s}$ from every TA event
as the positrons and radio waves originated from the same location at the same instant and both propagate with the speed of light. As  the typical time and spatial spreads of the TA events do not exceed $10 {\rm\,\mu s}$,   the  synchronization must be within the same temporal and spatial range. 

This section is organized as follows. In next subsection \ref{sect:E-field} we study the trajectory of the liberated positrons in   a typical electric field characterized by parameters (\ref{parameters1}). We analyze the total intensity and the angular distribution of the radio emission in subsection \ref{Intensity}, while  the spectral properties of the pulse are studied in subsection \ref{spectrum}. Finally, in subsection \ref{pulse} we study the properties of the   electric field's pulse   which can be detected  at the surface detector   area. 

\subsection{ Positron's acceleration in electric field under thunderstorm}\label{sect:E-field} 
Our goal here is to describe the dynamics of the liberated positrons which are emitted with initial energies in keV range according to Eq. (\ref{Delta_E}).
To simplify things we assume that the electric field ${\cal{E}} $ directed along $\mathbf{z}$-axis is uniform and characterized by parameters (\ref{parameters1}). 
In this case the solution is well known \cite{Landau} and can be described as follows. At the very initial moment of acceleration at $t\ll t_0$ the particle 
moves with non-relativistic velocities according to conventional formulae:
\be
\label{non-relativistic}
&&z(t)\simeq \frac{p_z^0}{m}t+\frac{|e|{\cal{E}} }{2m}t^2, ~~~~t\ll t_0\equiv \frac{mc}{|e|{\cal{E}} }  \nonumber\\
&&x_{\perp}(t)\simeq \frac{p_{\perp}^0}{m}t, ~~~  \frac{p_{\perp}^0}{m}\simeq \frac{p_z^0}{m}\simeq \frac{v_{\perp}}{c}  ,
\ee
where $p_z^0\sim p_{\perp}^0$ is the initial positron's momentum at the moment of  liberation from AQN. 
Numerically, this non-relativistic portion of the positron's journey is very short as $t_0\sim 10^{-8} \rm s$. It will be ignored    in  all our discussions which follow. 

The portion of the  positron's journey which plays the key role in our discussions is the motion with constant acceleration $e{\cal{E}}/m$  and the velocity     close to $c$. The corresponding motion is determined by the formulae: 
\be
\label{relativistic}
&&z(t)=  \frac{c}{|e|{\cal{E}} }\left[\sqrt{(p_{\perp}^0)^2+m^2c^2+(p_z^0+|e|{\cal{E}} t)^2}-\frac{E^0}{c}\right],  \nonumber\\
&&x_{\perp}(t)\simeq\frac{cp_{\perp}^0}{|e|{\cal{E}} }\ln \frac{2|e|{\cal{E}} t}{mc}, ~~~~t\gg t_0\equiv \frac{mc}{|e|{\cal{E}} } 
\ee
where $E^0\simeq (mc^2+\Delta E)$ is the total initial energy of positrons at the moment of liberation. 

The most important observation here is that the positrons   move with ultra-relativistic velocities for most of the time, and the displacement  in transverse direction  (with respect to the orientation of the electric field) is very modest such that the majority of the particles remain in  the system and continue their acceleration for the entire time interval $  \tau_{\cal E}\sim 0.3 \,\mu \rm s$ determined by parameter $l_a$ according to  (\ref{parameters1}). 
At this point most of the particles assume very high energy close to 10 MeV according to Eq. (\ref{E_MeV}) while   transverse momentum distribution is characterized by $(v_{\perp}/c)\lesssim 0.1$.  The elastic scattering of the positrons off the atmospheric molecules during this short  journey    plays very minor role
as the kinetic energy of  positrons  being  in keV range  at the  initial  instant  assumes   MeV energies  very quickly such that the corresponding cross section never becomes sufficiently large   to modify the positron's dynamics as computed above in vacuum.

\subsection{Intensity and the angular distribution}\label{Intensity}
Our next task is to present conventional formulae for the intensity and the angular distribution of the radio  emission due to the relativistic positrons (\ref{relativistic}) 
accelerating in electric field $\mathbf{{\cal{E}} } $ with parameters (\ref{parameters1}). As we discussed above the velocity  $\mathbf{v}$ of the positrons is mostly oriented along the electric field while $v_{\perp}$ is small. Therefore,  we assume that $\mathbf{{\cal{E}} } \parallel \mathbf{v}$ to simplify our formulae which follow. In this simplified geometry the angular distribution for the intensity assumes the following   form, see e.g. \cite{Landau,jackson}:
\be
\label{angular-distribution}
\frac{dI (t)}{d\Omega}\simeq \frac{N^2e^2 \mathbf{a}^2(t)}{4\pi c^3}\frac{\sin^2\theta}{(1-\frac{v}{c}\cos\theta)^5}, ~~~ \mathbf{a}\equiv \frac{e  \mathbf{{\cal{E}} }}{m\gamma^3},
\ee 
where $N$ is the number of coherent positrons participating in the radio wave emission, to be estimated below. The angle $\theta$ in this expression is defined as usual as the angle between the velocity $\mathbf{v}$ of the charged particle at the moment of emission  and the direction $\mathbf{n}$ of the observer, i.e. $\mathbf{v}\cdot \mathbf{n}=v\cos\theta$. Formula (\ref{angular-distribution}) explicitly shows that the emission   mostly occurs along $\mathbf{{\cal{E}} }$ direction   as $\mathbf{{\cal{E}} } \parallel \mathbf{v}$. This is precisely  the same direction where  burst events are recorded as shown on  Fig. \ref{geometry}. 

Integration over all angles leads to well known expression for the total intensity: 
\be
\label{intensity}
I(t)\simeq \frac{2N^2e^2 }{3c^3} \left[\mathbf{a}^2(t)\gamma^6\right] , ~~~ \gamma\equiv\frac{1}{\sqrt{(1- {v^2}/{c^2})}}
\ee 
Formula (\ref{intensity}) represents the well known feature of the emission that the intensity is strongly enhanced for ultra-relativistic particles, which is obviously the case for positrons accelerated up to 10 MeV energies according to Eq. (\ref{E_MeV}). This implies that the relativistic enhancement factor entering Eq. (\ref{intensity}) is enormous: $(E_{\rm exit}/m)^6\sim 10^8$.

\subsection{Spectral characteristics of the radio emission} \label{spectrum} 
The main goal of this subsection is to understand the spectral characteristics of the radio emission. It will also  allow to estimate the coherence factor $N$ entering Eqs. (\ref{angular-distribution}) and (\ref{intensity}) as this factor obviously depends on the frequency of radiation and  
corresponding  wave length. 

The starting point for these studies is spectral property of the electric field $\mathbf{E}_{\omega}$ emitted by the accelerating positrons, see e.g. \cite{Landau,jackson}:
\be
\label{spectral} 
\mathbf{E}_{\omega}=\int_{-\infty}^{+\infty}\mathbf{E}e^{i\omega t}dt, ~~ \mathbf{E}=\frac{Ne}{c^2R}\frac{\mathbf{n}\times\left((\mathbf{n}-\frac{\mathbf{v}}{c}\right)\times \mathbf{a})}{(1-\frac{\mathbf{n}\cdot \mathbf{v}}{c})^3},~~~ 
\ee 
where all quantities at the right hand side of Eq. (\ref{spectral}) must be computed as the retarded times $t'$:
\be
t'\approx t-\frac{R_0}{c}+\frac{\mathbf{n}\cdot\mathbf{v}t'}{c}~~ \rightarrow~~ t=t' \left(1-\frac{\mathbf{n}\cdot \mathbf{v}}{c}\right)+\frac{R_0}{c}, ~~~
\ee
where we assumed  that the positrons move with approximately constant time-independent velocity $|\mathbf{v}|\approx c$. This assumption allows us to represent the Fourier component $ \mathbf{E}_{\omega}$ in the following form
\be
\label{E_omega} 
\mathbf{E}_{\omega}=  \frac{e^{ikR_0}}{R_0} \left(\frac{Ne}{c^2}\right) \left(\frac{\omega}{\omega'}\right)^2\left[\mathbf{n}\times\left((\mathbf{n}-\frac{\mathbf{v}}{c})\times \mathbf{a}_{\omega'}\right)\right] ,~~~ 
\ee 
where $\omega'$ and  $\mathbf{a}_{\omega'}$ are defined as follows
\be
\label{omega'}
\omega'\equiv \omega \left(1-\frac{\mathbf{n}\cdot \mathbf{v}}{c}\right),  ~~~  \mathbf{a}_{\omega'}=\int_{-\infty}^{+\infty}\mathbf{a(t')}e^{i\omega' t'}dt'
\ee
which precisely the combination   entering the Fourier transform (\ref{spectral}).  

Using Fourier component  for $ \mathbf{E}_{\omega}$ as given by Eq. (\ref{E_omega}) and magnetic component  $ \mathbf{B}_{\omega} =i(\mathbf{k}_{\omega}\times \mathbf{E}_{\omega})$
one can  compute the   energy $dE_{\mathbf{n}\omega}$  emitted into  solid angle $d\Omega$  with frequency interval $d\omega$ \cite{Landau,jackson}:

\be
\label{E_omega1}
\frac{dE_{\mathbf{n}\omega}}{d\Omega d\omega}=   \left(\frac{N^2 e^2}{4\pi^2 c^3}\right)\left(\frac{\omega}{\omega'}\right)^4\left|\mathbf{n}\times\left((\mathbf{n}-\frac{\mathbf{v}}{c})\times \mathbf{a}_{\omega'}\right)\right|^2 . ~~~
\ee 
One can simplify this expression by assuming that the $\mathbf{a}\parallel \mathbf{v}$ as we previously discussed. Furthermore, one can carry out the integration over  $d \Omega$ by integrating over $d \omega'$ for a given $\omega$ as these two variables are related according to (\ref{omega'}). Indeed,
the emission is mostly concentrated along the cone with $\theta^2\approx (1-v^2/c^2)$. Furthermore, 
\be
\label{variables}
\omega'\equiv \omega \left(1-\frac{\mathbf{n}\cdot \mathbf{v}}{c}\right)\approx \frac{\omega}{2} \left[\frac{1}{\gamma^2}   + {\theta^2} \right],
\ee
where we expanded $\cos\theta\approx (1-\theta^2/2)$ and represented small factor $(1-v/c)\approx  {1}/{(2\gamma^2)}$  in terms of the conventional combination 
$\gamma$ which is valid  approximation for ultra-relativistic positrons with $|\mathbf{v}|\simeq c$. Now,  integration over angles $d \Omega$   can be replaced by integration over $d\omega'$   as these two variables are related  by (\ref{variables}). Therefore,  
\be
\label{variables-change}
d\Omega=2\pi \sin\theta d \theta\approx 2\pi\left(\frac{d\theta^2}{2} \right)\approx 2 \pi \left(\frac{d\omega'}{\omega} \right)
\ee
such that the  energy $dE_{\omega}$ emitted within the  frequency interval $d\omega$
assumes the form
\be
\label{E_omega2}
\frac{dE_{\omega}}{ d\omega}&=&   \left(\frac{N^2 e^2\omega^2}{2\pi c^3}\right)\int_{\frac{\omega}{2\gamma^2}}^{\infty}\frac{d\omega' |\mathbf{a}_{\omega'}|^2}{(\omega')^3}\left[2-\frac{\omega}{\omega' \gamma^2} \right] ,
\ee 
where the low limit of integration is determined by $\theta=0$ in relation (\ref{variables}).

The same expression can be thought as the angular distribution of the emission
\be
\label{angular-distribution1}
\frac{dE_{\omega}}{ d\omega}&=&   \left(\frac{N^2 e^2\gamma^6 }{2\pi c^3}\right) \frac{16|\mathbf{a}_{\omega'}|^2}{ (1+\gamma^2\theta^2)^3}\left[ \frac{\gamma^2\theta^2}{1+\gamma^2\theta^2} \right] \cdot \frac{d\Omega}{2\pi},~~~
\ee 
where $|\mathbf{a}_{\omega'}|$ should be  expressed in terms of $\theta$ according to relation (\ref{variables}).

Our next task is to model the acceleration $\mathbf{a}(t)$ for a typical thunderstorm electric field with parameters (\ref{parameters1}). The corresponding Fourier transform (\ref{omega'}) determines  $\mathbf{a}_{\omega'}$ which enters  expression for $ {dE_{\omega}}/{ d\omega}$ as given by Eq. (\ref{E_omega2}). Our simplest possible choice is as follows:
\be
\label{acceleration}
\mathbf{a}(t)\approx\frac{e  \mathbf{{\cal{E}} }}{\gamma^3 m}, ~~~ t\in (0, \tau_{{\cal{E}}} ), 
\ee   
while $\mathbf{a}(t)\approx 0$ for $t$ being outside of this interval. The corresponding expression  for   $\mathbf{a}_{\omega'}$ defined by (\ref{omega'}) assumes the form
\be
\label{acceleration'}
| \mathbf{a}_{\omega'}|^2=\frac{4e^2{\cal{E}}^2}{m^2\gamma^6\omega'^2}\sin^2(\frac{\omega' \tau_{{\cal{E}}}}{2}).
\ee
Now we can perform the integration $d\omega'$ in Eq. (\ref{E_omega2}).  For our simple model for acceleration  (\ref{acceleration}) the integral can be approximately  computed for small $(\omega' \tau_{{\cal{E}}})\ll 1$. 
This  leads us to the    order of magnitude estimate   for $ {dE_{\omega}}/{ d\omega}$:
\be
\label{E_omega3}
\frac{dE_{\omega}}{ d\omega}\approx   \left(\frac{N^2 e^2 }{2\pi c^3}\right)   \cdot \left(\frac{e^2{\cal{E}}^2\tau^2_{{\cal{E}}}}{m^2\gamma^6}\right) \cdot \left(\frac{4\gamma^4}{3}\right).
\ee 
The same expression can be derived from Eq. (\ref{angular-distribution1}) by integrating over the angles $d\Omega$. 

This spectral density holds as long as  $\omega$ is sufficiently small. To be more precise:
\be
\label{omega}
\omega     \lesssim   { \gamma^2}{\tau^{-1}_{{\cal{E}}}},
\ee
which represents the dominant contribution to the integral (\ref{E_omega2}) for small $\omega$. The emission with higher frequencies will be power-suppressed
as $\omega^{-2}$. 

Few comments are in order. The spectral density (\ref{E_omega3}) integrated over all   angles $d\Omega$ approximately constant and does not depend on $\omega$ as long as  
condition (\ref{omega}) is satisfied.  This is of course a well-known feature of the constant acceleration (\ref{acceleration}) which we used as a simplified model for the electric field. In reality, the electric field under thunderstorm obviously fluctuates in time and space. This will obviously modify the spectral features (\ref{E_omega3}). However, we expect that  a typical frequency of the emission as given by (\ref{omega}) will hold because  it is basically determined by the typical time scale of the problem $\tau_{{\cal{E}}}$ and expected relativistic factor $\gamma$ which is not very sensitive to the details of the fluctuating electric field  ${\cal{E}}$. 

The crucial observation here is that the typical frequency of emission is not simply given by the scale $\tau^{-1}_{{\cal{E}}}$ as one could naively expect based on dimensional arguments. Rather, the emission extends to  much broader region due to the large relativistic factor $\gamma^2$ as inequality (\ref{omega}) states. 

We postpone for the detail numerical estimates to Sect. \ref{numerics}. Now we want to make few simple numerical estimates supporting the main claim of this work that the frequency of emission is in the radio band $\nu\in (0.5-200)$ MHz.  Furthermore, the emission is mostly oriented along the same direction where TA bursts are recorded. Therefore, one should anticipate a strong synchronization  between the TA bursts and the radio pulses as both emission occurs at the same location at the same instant, and propagate  to the corresponding detectors with the same speed of light. 
These features are not very   sensitive to details  of the dynamics of the AQNs, nor specific features of  electric field under thunderstorm. Rather, all   these features  are inevitable  consequences  of the basic  AQN framework along with  pure geometrical properties of these observables.

With these comments in mind we can represent the  typical band  as follows 
\be
\label{nu}
\nu\lesssim \frac{{ \gamma^2}}{2\pi{\tau_{{\cal{E}}}}}\approx  200~ {\rm MHz},~~~~ \nu\equiv \frac{\omega}{2\pi}, ~~~ \gamma\approx 20,
\ee
which we   expect to hold irrespective of   specific features of the AQN model as it is entirely determined by well established 
typical   time scales  of the electric field under the thunderstorm (\ref{parameters1}). One could naively think that the typical frequency of emission could be extrapolated to very low $\nu$  as   the spectral density (\ref{E_omega3})   apparently does not depend on frequency. This is not quite correct conclusion though 
as we shall discuss in next Sect. \ref{numerics}.

\subsection{Radio pulse of the electric field}\label{pulse}
The goal of this section is to estimate the intensity of the  electric field 
(\ref{spectral})  at very large distances $R$ where it can be potentially detected. The orientation of  $\mathbf{E}$  field 
is determined by cross product  (\ref{spectral}) where one can    assume (for the simplicity of the numerical estimates) that $\mathbf{v}\parallel \mathbf{a}$, similar to our previous estimates for the spectral intensity of the emission (\ref{E_omega2}).
The absolute value for $|\mathbf{E}|$ at large distances  can be estimated as follows
\be
\label{E-field}
|\mathbf{E}|\approx \frac{Ne  |\mathbf{a}|  \theta}{c^2R} \left(\frac{\omega}{\omega'}\right)^3 \approx  \frac{Ne  |\mathbf{a}|  \theta}{c^2R}\left(\frac{2\gamma^2}{1+\gamma^2\theta^2}\right)^3,
\ee
where $R$ is the distance from the emission site  to the area  where the electric field $|\mathbf{E}|$ could be  recorded. It should not be confused with parameter $r$ which enters all formulae from Sect. \ref{TA-overview} and describes the distance from the emission site to the SD site where  the energetic  positrons could be detected. Numerically these parameters are similar, of course. However, the radio signal can be observed in a   different location from the local area where the positrons hit the TASD. 

This formula explicitly shows that the time duration of the radio pulse and its properties are unambiguously determined by the features of the electric field $\cal{E}$ under thunderstorm. For simple model (\ref{acceleration})  this implies that the radio pulse  lasts $\tau_{\cal{E}}$ while the absolute value of the pulse is estimated as
\exclude{
\be
|\mathbf{E} (\theta,t)| \approx  \frac{Ne    \theta}{c^2R}\left(\frac{2\gamma^2}{1+\gamma^2\theta^2}\right)^3 \left(\frac{e\mathbf{{\cal{E}} }(t)}{m \gamma^3}\right), \quad t\in (0, \tau_{{\cal{E}}} ).
\ee
}
\begin{equation}
\label{E-pulse}
|\mathbf{E} (\theta,t)| \approx  \frac{Ne    \theta}{c^2R}\left(\frac{2\gamma^2}{1+\gamma^2\theta^2}\right)^3 \left(\frac{e\mathbf{{\cal{E}} }(t)}{m \gamma^3}\right)\,, \quad t\in (0, \tau_{{\cal{E}}} )\,.
\end{equation}
It is important  to emphasize that the duration of the pulse $\tau_{{\cal{E}}} $ is not very sensitive  
to the details of the AQN model as it is entirely determined by well established 
typical   time scales  of the electric field under the thunderstorm (\ref{parameters1}). This feature is very  similar to our previous arguments regarding a typical frequency of the radio emission (\ref{nu}). At the same time the intensity of the pulse is highly sensitive to the details of the AQN model as it depends on the number of positrons $N$ participating in the emission. In this respect this feature of  insensitivity to any specific details of the AQN model for   the pulse duration  $\tau_{{\cal{E}}} $ and the frequency $\nu$ as given by (\ref{nu}) is similar to our studies of the TA burst events when such features as the ``curvature", the timing and the spatial spread of the bursts are not sensitive to the details of the AQN model, but entire determined by the geometry as reviewed  in Sect. \ref{TA-overview}.  
This should be contrasted with estimations \cite{Zhitnitsky:2020shd} of the intensity of the TA bursts which are highly  sensitive to specific   features of the model, which are hard to carry out in a quantitative manner.

\section{Numerical Estimates}\label{numerics}
Our goal here is to present some numerical results using parameters which had been used in our previous estimations \cite{Zhitnitsky:2020shd}  related to puzzling TASD bursts   as reviewed in Sect. \ref{TA-overview}.  One should emphasize from the very beginning that the corresponding estimates suffer from huge uncertainties, and should be considered as the order of magnitude estimates, at the very best. Our main arguments of the present work are not based on these
highly model dependent estimates. Rather, our main arguments are based on  essentially model-independent features of the radio emission such as 
typical frequency band (\ref{nu}) and strong synchronization between TASD burst events and the radio emission. Nevertheless,  we think that an order of magnitude estimates for the amplitude of the electric field as presented below could be useful as they demonstrate the consistency of the proposal.
The same estimates also demonstrate the principle feasibility to detect such radio signals.

\begin{figure}[ht]
	\centering
	\includegraphics[width=\linewidth]{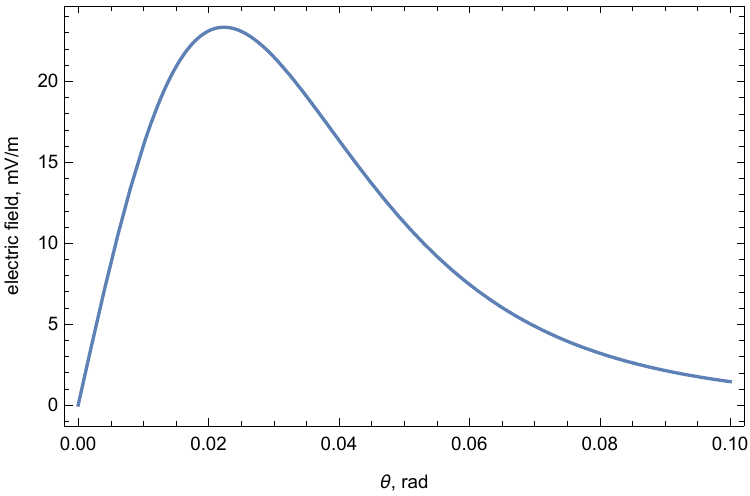}
	\caption{Strength of electric field (\ref{E-numerics}) induced under thunderstorm versus observation angle $\theta$. The parameters are chosen to be $N=10^9$, $\gamma=20$, ${\cal E}=1{\rm\,kV/cm}$.}
	\label{fig:thundercloud_E_theta}
\end{figure}

We start with numerical estimation of the strength of the electric field of the pulse (\ref{E-pulse}) in conventional units $\rm (mV/m)$:  
\be
&&|\mathbf{E} (\theta,t)| \approx  90 {\rm  \frac{mV}{m}}   \cdot \left[\frac{(\gamma\theta)}{(1+\gamma^2\theta^2)^3} \right] \nonumber \\
&&  \cdot   \left(\frac{\gamma}{20 } \right)^2\cdot \left(\frac{N}{10^9}\right)\cdot \left(\frac{\rm 10 ~ km}{R}\right), ~~~ ~~~~~~ t\in (0, \tau_{{\cal{E}}} ) \label{E-numerics}
\ee
where factor $N$ is the number of the coherent positrons participating in the emission, to be estimated below. 
The orientation of the electric field is  determined by the cross product as given by (\ref{spectral}), and it is approximately (up to small angle $\theta$) points along 
$\mathbf{a}$ which represents the direction of the   electric field under thundercloud at the instant of emission.  The temporal shape of the pulse (bipolar, unimodal or even more complicated form)  is also determined by the same fluctuating electric field $\cal{E}$ at the moment of emission. The intensity of the field is strongly peaks along  $\mathbf{n}$ with typical angle $\theta\gamma\lesssim 1$,  see Fig. \ref{fig:thundercloud_E_theta} for a precise angular distribution of $|\mathbf{E} (\theta)|$.

Our next task is the estimation of parameter $N$ entering Eq. (\ref{E-numerics}). 
Our  estimates will be   entirely  based on the observations, rather than on theoretical computations which inevitably suffer from huge uncertainties 
as reviewed in Sect. \ref{TA-overview}. Therefore, the logic of our estimation of parameter $N$ is as follows. Let us assume that the TASD bursts which have been recorded are indeed due to the positrons emitted by AQNs as proposed in \cite{Zhitnitsky:2020shd}.  
In this case the number of detected particles must be proportional to the density of particle detectors.  This number  is known for presently operating  507 SD detectors   with area $3 \rm  m^2$ each. These detectors are 
covering $680 ~\rm km^2$ total area. The number of detected particles  must be also proportional to the area $\sim \pi (r\Delta\alpha)^2$ where particles had been recorded by TASD during   a single event within the burst.  Therefore, the number of positrons $N$ at the emission site  can be estimated from the following relation:   
\exclude{
\be
\label{N}
N_{\rm positrons}^{\rm detected}[\Delta s]\approx N  \left[ \rm\frac{   507\cdot 3 ~m^2 }{680 ~km^2} \right] \left[ \la \exp\left(-\frac{r}{\lambda}\right)\ra \right]  \left[ \pi (r\Delta \alpha)^2\right]~~~~~~
\ee
}
\begin{equation}
\label{N}
N_{\rm positrons}^{\rm detected}[\Delta s]\approx N  \left[ \rm\frac{   507\cdot 3 ~m^2 }{680 ~km^2} \right] \left[ \la \exp\left(-\frac{r}{\lambda}\right)\ra \right]  \left[ \pi (r\Delta \alpha)^2\right]
\end{equation}
In our estimate (\ref{N})  the number of detected particles being  recorded over the surface  area $\pi (\Delta s)^2$ [left hand side of (\ref{N})] varies for different events within a single  burst, and  it  normally varies in the range $(2-6)\cdot10^2 $. In our estimate (\ref{N}) we inserted the   suppression factor 
$\la \exp(-r/\lambda)\ra\sim 0.1$  introduced in \cite{Zhitnitsky:2020shd}  to account for some attenuation of  the positrons  travelling a distance $r\sim$ 10  km   or so when the    mean free path $\lambda$ for positrons with few MeV energy  is order of kilometre at the sea level and several  kilometres at higher  altitudes.   One should also note that $\pi (r\Delta\alpha)^2$ on the right hand side of Eq. (\ref{N}) is not identically the surface area $\sim \pi(\Delta s)^2$ where particles are being   recorded. These areas are directly related to each other through angle $\alpha$  according to Eqs. (\ref{spread2}), see Fig. \ref{geometry} for notations.

The relation (\ref{N}) suggests that number $N$ entering Eq.  (\ref{E-numerics}) 
for   $(r \Delta \alpha) \sim 1 ~\rm  km$ (which corresponds to a typically observed spatial spread   for the burst events)   can  be estimated as  
\be
\label{N1}
N\approx (0.3-1)\cdot 10^9  ~~~ {\rm for} ~~~ (r \Delta \alpha) \sim 1 ~\rm  km.
\ee
This estimate for $N$  corresponds to the amplitude of the electric field (\ref{E-numerics}) on the level  $20\rm (mV/m)$    measured at distance $R\simeq 10$ km
from the source of the emission within the  angle range   $(\theta \gamma)\lesssim 1$ where the most of the radio emission occurs.

\exclude{We use two complimentary approaches to make the corresponding estimates. First one  is based on a pure theoretical consideration by computing the $T$-dependent  ionization  charge $Q (T)$ of the AQN presented  in ref.  \cite{Zhitnitsky:2020shd}, including (\ref{capture1}) as overviewed in Sect. \ref{AQN}. The second estimate is based on drastically different   approach when one counts the required number of positrons at the moment of emission to match the recorded by TASD  number of particles for a single event within  the burst.  
	
	We start with pure theoretical estimate when $Q (T)\approx 10^{11} $ for $T\approx 10$ keV as estimated in \cite{Zhitnitsky:2020shd}.
	This number should be multiplied by factor $\eta(T)\simeq 0.1$ which accounts for the portion of the liberated positrons.  
	One should emphasize that there are many uncertainties in these estimates which are beyond of the theoretical control as they depend on interaction of fast moving (with large Mach number) AQN under the thunderclouds, when the turbulence, shock waves and other complicated phenomena occur.
	This ``theoretical"  approach leads to the following  estimate: $N_{\rm theory}\simeq \eta Q\sim 10^{10}$. 
}

Our last task in this section is   establishing  the lower frequency bound for the radio emission.  The spectral density (\ref{E_omega3}) which is approximately a constant, naively suggests that the spectrum extends to arbitrary low frequencies. In fact, it is a premature conclusion as formula  (\ref{E_omega3}) was derived assuming a constant acceleration. In reality, a typical acceleration time is determined by (\ref{parameters1}) such that one should expect that
$ \nu\gtrsim  (2\pi{\tau_{{\cal{E}}}})^{-1}\approx  0.5~{\rm MHz}$.  Therefore,   the AQN-induced   $\sim 0.3\mu$s pulse   represents a    radio emission  in the   bandwidth    $\nu\in (0.5-200) \rm ~MHz$  with  the amplitude of the corresponding electric field of the order   $|\mathbf{E} (\theta,t)| \sim20\rm (mV/m)$  
at distance $R\sim 10 \rm~ km$ as estimated above.  
\exclude{
	\begin{figure}[ht]
		\centering
		\includegraphics[width=1\linewidth]{thundercloud_spectrum_normalized.pdf}
		\caption{Normalized spectrum of (\ref{E_omega2}) in terms of $\omega \tau_{\cal E}/\gamma^2$. The full width at half maximum gives $\omega\tau_{\cal E}/\gamma^2\lesssim3$.}
		\label{fig:thundercloud_spectrum_normalized}
	\end{figure}
}
Important feature of this spectrum is that it is very broad. Indeed, the intensity of the radiation  changes by factor  two or so when  the  frequency varies by two orders  of magnitude. Another comment which follows from  Fig. \ref{fig:thundercloud_spectrum_MHz} is that the portion of the positron's energy being converted into the radio waves is very  tiny as initial positrons energy can be estimated as $10 N~ {\rm MeV}$ when $N$ is estimated in Eq. (\ref{N1}). This implies that the initial energy is    many orders of magnitude greater than the radio wave energy shown on
Fig. \ref{fig:thundercloud_spectrum_MHz}.

\begin{figure}[ht]
	\centering
	\includegraphics[width=\linewidth]{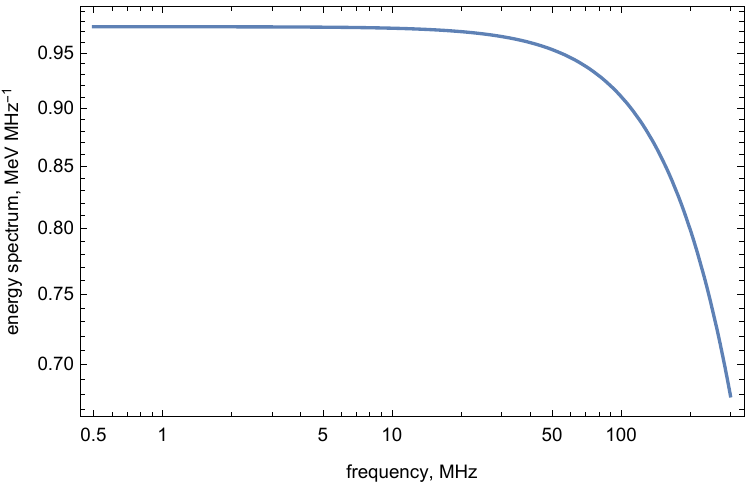}
	\caption{Spectrum (\ref{E_omega2}) in terms of frequency $\nu$, thunderstorm radiation.  The parameters are chosen to be $N=10^9$, $\gamma=20$, ${\cal E}=1{\rm\,kV/cm}$.}
	\label{fig:thundercloud_spectrum_MHz}
\end{figure}

The crucial point of our analysis can be formulated  as follows. This radio pulse must be synchronized with  an  event within  TA burst within $10\,\mu$s because the TA burst and the radio emission are originated from the same location 
emitted at the same instant and both signals propagate with the speed of light along  the  same direction to be detected within $\sim \rm km^2$ area. 

It is important  to emphasize that such strong synchronization must hold even if the TA bursts are not related to any lightning events and can be easily discriminated from much stronger radio emission which always accompany   lightnings. Therefore, such  synchronization between the radio pulses and the TA bursts, if observed, can be easily distinguished  from any other possible correlation between radio emission  and conventional lightning events, see Sect. \ref{mimic} for the details.

\section{Geosynchrotron radiation}\label{sec:Geosynchrotron radiation}
Another significant radio emission is the geosynchrotron radiation which is due to 
presence of the Earth's magnetic field   ${\cal B}\sim0.5{\rm\,gauss}$. This effect becomes important  near the end of radio emission under thunderstorm. At this instant    all the positrons are ultrarelativistic ($\gamma\sim 20$)  but they are  no longer accelerating  as they already left the region of the  external electric field $\cal{E}$. Trajectory of positrons are helical in presence of the magnetic field of Earth ${\cal B}\sim0.5{\rm\,gauss}$. The acceleration in this case is perpendicular to the direction of velocity $\mathbf{v}\perp\mathbf{a}_{\cal B}$, so it is convenient to set up the following coordinate with respect to the direction of observer $\mathbf{n}$:
\be
\hat{\mathbf{v}}\cdot\mathbf{n}
=\cos\theta\,,\quad
\hat{\mathbf{a}}_{\cal B}\cdot\mathbf{n}
=\sin\theta\cos\phi
\ee
Radius of the helical trajectory can be estimated as 
\exclude{
\be
&\rho
\approx\frac{c^2}{|\mathbf{a}_{\cal B}|}
\approx\frac{0.7{\rm\,km}}{\sin\theta_{\cal B}}\left(\frac{\gamma}{20}\right)
\left(\frac{0.5{\rm\,gauss}}{{\cal B}}\right)\,,\quad
&|\mathbf{a}_{\cal B}|
\approx\frac{e{\cal B}c\sin\theta_{\cal B}}{\gamma m}\,,
\ee
}
\begin{equation}
\begin{aligned}
&\rho
\approx\frac{c^2}{|\mathbf{a}_{\cal B}|}
\approx\frac{0.7{\rm\,km}}{\sin\theta_{\cal B}}\left(\frac{\gamma}{20}\right)
\left(\frac{0.5{\rm\,gauss}}{{\cal B}}\right)\,,\\
&|\mathbf{a}_{\cal B}|
\approx\frac{e{\cal B}c\sin\theta_{\cal B}}{\gamma m}\,,
\end{aligned}
\end{equation}
where $\theta_{\cal B}$ is defined to be the angle between direction of ${\cal B}$ and the velocity of positrons $\mathbf{v}$. 

Radiation from an ultrarelativistic particle has a narrow emission angle $\theta\lesssim\gamma^{-1}$, similar to our previous discussions. However, the pulse of  radiation for geosynchrotron radiation  is much shorter    \cite{Landau,jackson}:
\be
\tau_{\cal B}
\approx\frac{\rho}{2\gamma^3c}
=\frac{0.14{\rm\,ns} }{\sin\theta_{\cal B}}
\left(\frac{20 }{\gamma}\right)^2
\left(\frac{0.5{\rm\,gauss}}{{\cal B}}\right) 
\ee
This interval is much shorter than the time scale $\tau_{\cal{E}}\sim 0.3{\rm\,\mu s}$ discussed previously 
due to extra  suppression factor $\sim\gamma^{-3}$.  
The energy spectrum integrated over angle $d\Omega$ is well known \cite{Schwinger:1949,Landau,jackson}:

\exclude{
\be
&\frac{d E_\omega}{d\omega}
=\frac{c\tau_{\cal B}}{2\pi\rho}\frac{\sqrt{3}N^2e^2}{c}\gamma
\frac{\omega}{\omega_c}\int_{\omega/\omega_c}^{\infty}K_{5/3}(x)dx\,, \quad
&\omega_c
=\frac{3}{2}\gamma^3\left(\frac{c}{\rho}\right) \simeq \frac{3}{4\tau_{\cal B}}\,,
\ee
}
\begin{equation}
\label{eq:E_omega geosyn}
\begin{aligned}
&\frac{d E_\omega}{d\omega}
=\frac{c\tau_{\cal B}}{2\pi\rho}\frac{\sqrt{3}N^2e^2}{c}\gamma
\frac{\omega}{\omega_c}\int_{\omega/\omega_c}^{\infty}K_{5/3}(x)dx\,, \\
&\omega_c
=\frac{3}{2}\gamma^3\left(\frac{c}{\rho}\right) \simeq \frac{3}{4\tau_{\cal B}}\,,
\end{aligned}
\end{equation}
where $K_{5/3}(x)$ is the modified Bessel function, and $\omega_c$ is the critical angular frequency beyond which radiation becomes negligible. \ref{eq:E_omega geosyn}

The electric field of the geosynchrotron radiation follows from Eq. (\ref{spectral}):
\exclude{
\be
|\mathbf{E}|
&=\frac{Ne|\mathbf{a}_{\cal B}|}{c^2R}\left(\frac{\omega}{\omega'}\right)^2
\sqrt{
	1
	-\frac{(\mathbf{n}\cdot\hat{\mathbf{a}}_{\cal B})^2}{\gamma^2}
	\left(\frac{\omega}{\omega'}\right)^2
}  \nonumber \\ 
&\approx\frac{Ne}{c^2R}
\left(\frac{e{\cal B}c\sin\theta_{\cal B}}{\gamma m}\right)
\left(\frac{2\gamma^2}{1+\gamma^2\theta^2}\right)^2 
\sqrt{
	1-\left(\frac{2\gamma\theta\cos\phi}{1+\gamma^2\theta^2}\right)^2
}\,,  
\ee
}
\begin{equation}
\label{eq:geosyn E}
\begin{aligned}
|\mathbf{E}|
&=\frac{Ne|\mathbf{a}_{\cal B}|}{c^2R}\left(\frac{\omega}{\omega'}\right)^2
\sqrt{
	1
	-\frac{(\mathbf{n}\cdot\hat{\mathbf{a}}_{\cal B})^2}{\gamma^2}
	\left(\frac{\omega}{\omega'}\right)^2
}  \\ 
&\approx
\frac{Ne^2{\cal B}\sin\theta_{\cal B}}{\gamma mcR}
\left(\frac{2\gamma^2}{1+\gamma^2\theta^2}\right)^2 
\sqrt{
	1-\left(\frac{2\gamma\theta\cos\phi}{1+\gamma^2\theta^2}\right)^2
}\,,
\end{aligned}
\end{equation}
The estimation of electric field strength holds as long as the positron current remains coherent. The coherent length $l_{\rm coh}$ can be estimated in what follows. The dispersion in velocity is $\delta v\sim v_\perp\lesssim0.1c$ from Eqs. (\ref{spread2}). The condition of coherence requires the dispersion length $\delta l$ to be much less than the radius $\rho$ of trajectory:
\begin{equation}
\delta l
\approx\frac{\delta v}{c}l_{\rm coh}
\sim\frac{v_\perp}{c}l_{\rm coh}
\ll\rho\,.
\end{equation}
It implies $l_{\rm coh}\sim\rho$ that is insensitive to details of local parameters such as $\gamma$ and ${\cal B}$, and the geosynchrotron radiation is only significant within the first cycle of rotation.

\section{Numerical estimates  on geosynchrontron radiation}\label{B-numerics}
\begin{figure}[ht]
	\centering
	\includegraphics[width=\linewidth]{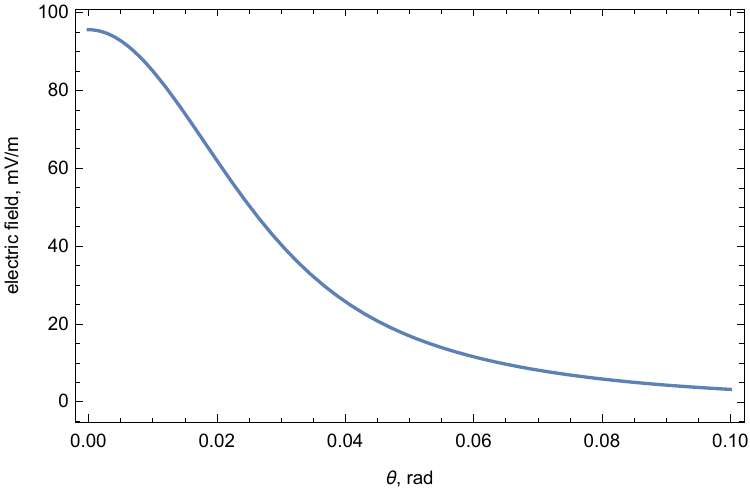}
	\caption{Strength of electric field (\ref{eq:geosyn E}) induced by geosynchrotron radiation versus obervation angle $\theta$. The parameters are chosen to be $N=10^9$, $\gamma=20$, ${\cal B}=0.5{\rm\,gauss}$, $\theta_{\cal B}=\phi=\pi/4$.}
	\label{fig:geosyn_E_theta}
\end{figure}

\begin{figure}[ht]
	\centering
	\includegraphics[width=\linewidth]{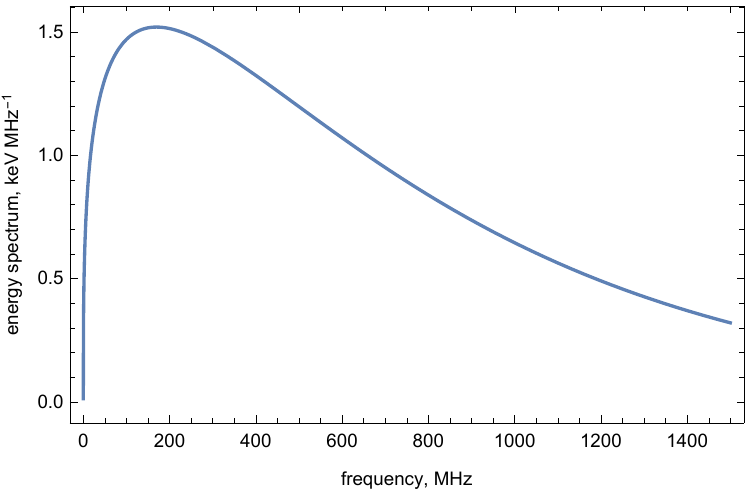}
	\caption{Spectrum (\ref{eq:E_omega geosyn}) in terms of frequency $\nu$, geosynchrotron radiation.  The parameters are chosen to be $N=10^9$, $\gamma=20$, ${\cal B}=0.5{\rm\,gauss}$, $\theta_{\cal B}=\pi/4$}
	\label{fig:geosyn_spectrum_MHz}
\end{figure}

Similar to Sect. \ref{numerics}, we make numerical estimation for the electric field strength and the frequency bandwidth. Numerical estimation for the strength of the electric field of pulse Eq. (\ref{eq:geosyn E}) in conventional units (mV/m) is given as:
\exclude{  
\be
|\mathbf{E}|
&\approx100 {\rm\,\frac{mV}{m}}\cdot
\frac{\sin\theta_{\cal B}}{(1+\gamma^2\theta^2)^2}
\sqrt{
	1-\left(\frac{2\gamma\theta\cos\phi}{1+\gamma^2\theta^2}\right)^2
} \nonumber\\
&\quad\times\left(\frac{\gamma}{10}\right)^3
\left(\frac{N}{10^9}\right)
\left(\frac{10{\rm\,km}}{R}\right)
\left(\frac{{\cal B}}{0.5{\rm\,gauss}}\right)\,,
\ee
}
\begin{equation}
\begin{aligned}
|\mathbf{E}|
&\approx100 {\rm\,\frac{mV}{m}}\cdot
\frac{\sin\theta_{\cal B}}{(1+\gamma^2\theta^2)^2}
\sqrt{
	1-\left(\frac{2\gamma\theta\cos\phi}{1+\gamma^2\theta^2}\right)^2
} \nonumber\\
&\quad\times\left(\frac{\gamma}{10}\right)^3
\left(\frac{N}{10^9}\right)
\left(\frac{10{\rm\,km}}{R}\right)
\left(\frac{{\cal B}}{0.5{\rm\,gauss}}\right)\,,
\end{aligned}
\end{equation}
see Fig. \ref{fig:geosyn_E_theta} for a precise angular distribution with specific parameters.
The upper limit of frequency band is determined by the critical frequency $\omega_c$: 
\begin{equation}
\nu
\lesssim\frac{\omega_c}{2\pi}
=\frac{3}{8\pi\tau_{\cal B}}
\approx600 {\rm\,MHz}\,,\quad
(\theta_{\cal{B}}=\frac{\pi}{4})\,,
\end{equation}
and the lower frequency cutoff is given by the fundamental frequency: 
\begin{equation}
\nu\gtrsim \frac{c}{2\pi\rho}\approx {50{\rm\,kHz}}\,,\quad
(\theta_{\cal{B}}=\frac{\pi}{4}).
\end{equation}
Therefore, the geosynchrotron induces a  radio pulse with time duration $\sim0.14{\rm\,ns}$  in the bandwidth $\nu\in(0.05-600){\rm\,MHz}$  (see Fig. \ref{fig:geosyn_spectrum_MHz})  and the amplitude of the corresponding electric field is of order $\sim 50{\rm\,mV/m}$ at distance $R\sim10{\rm\,km}$. 

\exclude{
	\begin{figure}[ht]
		\centering
		\includegraphics[width=1\linewidth]{geosyn_spectrum_MHz_loglog.pdf}
		\caption{\Red{log-log plot for illustrative purpose.}}
	\end{figure}
}

While numerical values of the frequency bandwidth $\nu\sim (0.1-10^2)~ \rm MHz$  and the maximal strength of electric field of the radio wave [$|\mathbf{E}|\sim (20 -50)~\rm mV/m$]   are similar in both cases    of radio pulses (which we term  as $\cal{E}$-type and $\cal{B}$-type correspondingly) there is an important    distinct feature, which is manifested in dramatically different   time durations in these two cases. To be more precise,    a $\cal{B}$-type pulse induced by geosynchrotron radiation ($\tau_{\cal B}\sim0.14{\rm\,ns}$)  is much shorter than $\cal{E}$-type pulse  with  ($\tau_{\cal E}\sim0.3{\rm\,\mu s}$).

\exclude{
	because $\tau_{\cal B}$ has an additional suppression factor $\gamma^{-2}$ while $\tau_{\cal E}$ is insensitive to $\gamma$. The $\gamma^{-2}$ suppression is a generic feature of radiation from an ultrarelativisitic source in nonlinear motion due to its narrow emission angle $\theta\lesssim\gamma^{-1}$. The absence of suppression factor for radiation under thunderstorm is a special case as the particle is linearly accelerated and has a fix emission direction toward the observer during the whole process. 
	
	Second, the strength of electric field $|\mathbf{E}|$ has a different proportionality to the boost factor $\gamma$ in the two cases. $|\mathbf{E}|$ is more sensitive to $\gamma$ for radio pulses induced by geosynchrotron ($\propto\gamma^{3}$) than the one induced under thunderstorm ($\propto \gamma^2$) because radiated power in synchrotron radiation is larger than parallel acceleration by a factor of $\gamma^2$. As $|\mathbf{E}|$ is proportional to square root of the emission power, this gives the electric field of geosynchrotron radiation an additional factor of $\gamma$ comparing to the one induced by coherent electric field under thunderstorm. 
	
	Comparing to the one under thunderstorm, radio pulses induced geosynchrotron have a much shorter time duration below nanosecond, but the signal strength $|\mathbf{E}|$ can be potentially more prominent by an order of magnitude in case of more energetic positrons $\gamma\gtrsim10$.
}

This distinct feature is  in fact  an  absolutely crucial element for possible potential observations of the  emission.     This difference between $\tau_{\cal B}$  and $\tau_{\cal E}$  is the main reason for dramatically different
energy injection which is measured in units $\rm MeV\cdot MHz^{-1}$ in case of $\cal{E}$-type emission to be contrasted with much lower $\rm keV\cdot MHz^{-1}$ scale in case of $\cal{B}$-type
emission, see Figs. \ref{fig:thundercloud_spectrum_MHz} and \ref{fig:geosyn_spectrum_MHz} correspondingly.

Therefore, the total   energy of the radio emission   integrated over the pulse  time is much greater for the $\cal{E}$-type radiation in comparison with $\cal{B}$-type radiation.  Furthermore, the required resolution to detect the very short $\cal{B}$-type  pulse with  $\tau_{\cal B} \sim \rm ns$  
requires  much more demanding instruments than recording  of  the conventional $\cal{E}$-type   pulses with duration time of order $\tau_{\cal E}\sim {\rm \mu s}$.
These arguments strongly suggest that $\cal{E}$-type radiation plays the dominant role in emission processes. Therefore, any future    studies (including possible observations) should be    focused   on $\cal{E}$-type radiation analyzed in Sect. \ref{numerics}.

\section{Can conventional physics be responsible for proposed effects?}\label{mimic}
The question we address in this section can be formulated as follows. It has been known for quite  sometime that the thunderstorm lightnings are always accompanied by the  radio   emissions, see e.g. reviews \cite{Gurevich_2001,DWYER2014147}.
Can these  lightning-induced radio pulses be responsible for the  effects studied in this work?
Before we answer this question we first recall \cite{Zhitnitsky:2020shd} the basic features of the TA bursts interpreted as   the AQN events, see item 1.
Then, in item 2 we overview    the main features of the radio emission which always accompanies   the thunderstorm lightning events. After that in item 3 we explain the dramatic qualitative differences between the radio  pulses   associated with thunderstorm lightnings and the radio pulses studied in this work.

{\bf1.}   As we previously mentioned, lightnings are correlated with AQN events when an AQN hits the region of the thunderclouds and potentially may initiate lightnings serving as a trigger \cite{Zhitnitsky:2020shd}. The mechanism is similar to the role of conventional CRs that are considered to be one of the dominant triggers of lightnings, see footnote \ref{trigger} with related comments and references. This is the basic reason why the  bursts being interpreted as random  AQN events are nevertheless  associated with the lightnings.

Further to this point, it is important to emphasize that  in the AQN framework the direction of the electric field $\cal{E}$ characterized by (\ref{parameters1}) (also shown on Fig. \ref{geometry}) and the direction of the lightning current are not correlated.
This is because the AQN event may serve as a  trigger   for the lightning strike.
The trigger  should not be confused with  the initial stepped lightning leader which determines the direction of the   current of the lightning being developed.
In different words, the direction of the observed current of a lightning does not determine the sign of the fluctuating electric field  $\cal{E}$ which itself defines the direction of the positrons to be recorded as TASD unusual events.

To rephrase it, an AQN plays the dual role when it propagates in  the thunderclouds:  it emits the very energetic  positrons, and it also triggers the lightning, similar to conventional CR, see footnote \ref{trigger} with relevant references. The former phenomenon is recorded as a TASD burst, while the latter one explains the observed correlation between the bursts and lightnings. 

As a next remark: as we mentioned, all mysterious bursts occur at the very initial moment of the lightning flashes   or even earlier than lightning. Some bursts are not related to the lightning flashes at all (but occur under the thunderclouds). In original interpretation   \cite{Abbasi:2017rvx} this feature is explained  in terms of  the initial stepped lightning
leaders propagating to the ground on a time scale of ten or tens of
milliseconds.    In the AQN framework this feature is  also automatically satisfied as the AQNs can serve as the triggers of lightning flashes, which may or may not initiate the lightnings. The particles which are recorded as TASD events are produced directly by  AQN itself  at this moment, not at the  initial stage of the strike   determined by the lightning leader.

\exclude{We also overview in subsection \ref{CR-induced}
	the properties of the radio emission associated with CR air showers   \cite{Huege:2003up,HUEGE2005116}. We explain the qualitative distinctions   
	between the generic features of the   CR-induced radio emission  and the radio pulses studied in this work. Our main conclusion is that the known radio emissions can be easily  discriminated from  the AQN-induced radio pulses studied in this work.
}


{\bf 2.}   Having discussed  the basic features of the TASD events within AQN framework we now turn to our next item on radio emission features as a result of   lightnings.  The correlation between lightning and radio emission during thunderstorms   is well known and well documented   generic feature of the  the  lightning discharges, see   \cite{GUREVICH2003228}  with large number of references on  observations. In particular, the     lightning discharges are  characterized by a very large number of radio pulses which last in total for about 1 second.  Each pulse is characterized by full width $(0.2-0.3)\mu {\rm s}$ with electric field strength   which could be as large as  $ |\mathbf{E}|\sim10^3  \rm ~mV/m$. Most of the pulses, though,  show  the strength of the electric field in the   
$|\mathbf{E}|\sim (100-200) \rm ~mV/m$ range.  Another important  feature of the radio emission: the gaps between pulses are in the range $ (10-10^2) \mu {\rm s}$. Therefore, total number of pulses could be very large $\gtrsim 10^3$ during a single lightning event.  Finally, the typical frequency of the radiation is strongly peaked in few MHz bands, while it completely diminishes for $\nu\gtrsim 10 \rm ~MHz$.

{\bf 3.} These features must be contrasted with  the AQN-induced radio pulses studied in this work. While a typical duration of an individual  pulse $\tau_{\cal E}$  and the strength of the corresponding electric field  $|\mathbf{E}|$  assume similar orders of magnitude in both cases, some other key characteristics are dramatically different.  It allows an easy discrimination between these two cases. 

Indeed, the number of clustered  radio pulses associated with the mysterious bursts  must be very few (it ought to be 3+ corresponding to the number of clustered events in a single burst). It should be    contrasted with $\sim 10^3$ in case of the 
lightning-induced radio pulses. The total durations of the radio emissions are  also very different for these two cases. 

The most important distinct feature which discriminates the different sources of the emission is that the frequency bands of the radiation are dramatically different in these two cases. The lightning-induced radio emission is strongly peaked in few MHz bands, while AQN-induced radio pulse is characterized by the flat spectrum with $\nu\lesssim 200 \rm ~MHz$ according to (\ref{nu}). The basic reason for  this dramatic difference is that the electric current responsible for the lightning is represented by the particles with $\gamma\sim1 $ while for the AQN-induced case the positrons are characterized by $\gamma\sim 20$. This difference is translated into dramatic modification of the frequency bands     according to (\ref{nu})   though  the typical time scale for an individual lightning-induced pulse  $\sim (0.2-0.3)\mu {\rm s}$ assumes the same order of magnitude as $\tau_{\cal E}\sim 0.3\mu {\rm s}$.

Next,   we want to mention that there  are some   dramatic differences between the radio pulses induced by conventional CR showers \cite{Huege:2003up,HUEGE2005116} and the AQN-induced radio pulses. These differences have been discussed in details in \cite{Liang:2021rnv}.  
Here we want to mention that these dramatic differences are related to very different structures of the showers.  In case of  conventional CR  the radio emission is  based on the picture  when a  CR air shower is   characterized by  the ``pancake"  of particles and its central axis.
The density of the particles strongly depend on the distance to the central axis such that  the spectral density of the radio pulse is highly sensitive to the width of the ``pancake", which becomes much thicker further away from the axis.   It is very different  from the AQN-induced radio signal as the notions of the  shower axis  and the ``pancake" do not exist in our case such that  the density of the particles is approximately the same irrespective of   the distance to the central axis. 

Additionally, conventional CRs are well studied even being heavily distorted by electric field in thundercloud, and in fact it is nowadays an efficient probe to study the electric field in thunderstorms by analysing the radio emission \cite{Schellart:2015kga}\footnote{Another way to study the electric field under the thunderstorm is to study distortion of the muon's propagation in the background of electric field \cite{Hariharan:2019fai}.}. The dramatic difference with the radio pulse studied in the present work is that  the radio emission due to the CR is strongly correlated with the Earth's magnetic field, in contrast with dominant $\cal{E}$-type  emission studied in the present work. As a result the width of the corresponding signal is in nanosecond  range, in contrast with the AQN induced pulse when the width is   $\sim 0.3 \mu {\rm s}$. Furthermore, the polarization pattern for the pulses related to CRs are dramatically different from the AQN-induced emission.     Therefore, there should be no technical difficulty to distinguish an AQN-induced radio pulse from a conventional CR under the thunderstorm. The corresponding studies could  support or refute our proposal.

There are many other distinct features between  the AQN and conventional CR air showers. In particular, the AQNs  mostly emit photons in X ray bands, such that a signal cannot  be observed by a fluorescence detector
which is designed to detect the visible  and UV  light\footnote{In fact, it was precisely the argument presented in \cite{Budker:2020mqk} why the all sky camera  has not observed any signal from a powerful meteor-like event detected by the infra-sound  dedicated instrument.}.

Next, we note that gamma rays can be produced by an avalanche of relativistic runaway electrons during initiation of lightning \cite{Dwyer:2003afl}. Bursts of gamma rays initiated in the Earth's atmosphere, commonly referred to as terrestrial gamma ray flashes, can be generated in a lightning leader system \cite{Pasko:2014emo,Abbasi:2018grs}. The terrestrial gamma ray flashes can trigger atmospheric photonuclear reactions that produce neutrons and positrons \cite{Enoto:2017lpx}. Evidently these gamma rays and their byproducts do not interfere with our proposed radio detection, as they are excessively more energetic and only appear at frequency band above EHz.  
Another important point here is that this powerful emission occurs in a latter stage of a lightning, while an AQN-induced event appears before or at the initial stage as it serves as a potential trigger of a lightning strike.

Based on these dramatic differences  in frequencies and   timings of the radio emissions   we conclude that the signals  due to the thunderstorm lightning events and conventional CR air showers can be easily  discriminated from  the AQN-induced radio pulses studied in this work.   Therefore, we suggest to study the corresponding radio signal to support or refute our proposal to interpret the TA bursts as the AQN-induced events. 

Lastly, one can assume  that the source of the TASD bursts is entirely due to the complicated and not well understood physics of the lightning strikes  (which may or may not be recorded by the system on the surface). In addition to many problems with explanation of the observed intensity, timing, clustering features, and geometry, there is an additional problem     of  the very low observed event rate (which is 10)    in comparison with recorded $\sim 10^4$ number of lightnings in the same area. At the same time the estimated event rate \cite{Zhitnitsky:2020shd} based on assumption  that the observed TASD bursts is a result of the AQN-induced events is consistent with observations. 

To conclude: we are not aware of any studies which could explain the   observed intensity, timing, clustering features, and geometry  of the signal in form of the mysterious bursts being produced during initial stage of the lightning as recorded, while proposal \cite{Zhitnitsky:2020shd} naturally  explains all these features within AQN framework. 
\exclude{In fact, it is very hard to imagine  how to produce simultaneously  at the very initial stage of the lightning processes a large number of  energetic coherent particles $N\simeq (10^8-10^9)$  which would be moving along the same direction for few kilometers to generate the observed TA signal.}

\section{Conclusion}\label{conclusion}
As we stated in the Introduction the main goal of the present work is to test the proposal   \cite{Zhitnitsky:2020shd}
   by searching  for  the     radio signals in  frequency band $\nu\in (0.5-200)$ MHz  which must be synchronized with the TA bursts.
   Such test would unambiguously support or refute the proposal (interpreting TA bursts as the AQN events) as the radio signals due to the AQN annihilation events can be easily discriminated from conventional radio pulses which always accompany the thunderstorm.
   This is precisely the goal of this work: we want  to eliminate all common  objections which essentially state that a thunderstorm is very  complicated system\footnote{For example, the question on what initiates and triggers a runaway breakdown avalanche still remains a matter of debates, in spite of many years of studies, see also footnote \ref{trigger}.} such that everything is possible, including TA bursts as a result of flashes. The proposal of the present work is to study   the     radio signals in  frequency band $\nu\in (0.5-200)$ MHz. The corresponding results  would unambiguously 
   answer these rhetoric questions, see Sect.  \ref{mimic} with more comments on this. 
   
  Now we summarize the results of our studies. 
As we argued at the very end of previous Sect. \ref{B-numerics} the $\cal{E}$-type radiation plays the dominant role in radio emission. The corresponding results have been summarized at the very end of Sect.  \ref{numerics}
where it has been argued    that the radio pulse 
can be recorded  if  proper instruments are designed and built at the TASD site.  The pulse   must be synchronized    with TASD bursts.    Such a signal  cannot be confused with any other spurious and noise signals as a result of this synchronization. 
Therefore, observing (not observing) such  synchronized   signals  can confirm, substantiate or refute our proposal.
We further argued in Sect. \ref{mimic} that the AQN-induced radio signals can be easily discriminated from the  pulses generated by the thunderstorm lightning events.

One should comment here that the    strength   (\ref{E-numerics}) of the amplitude  $|\mathbf{E}|$ on the level $20\rm~ mV/m$    with the duration  time    of the radio pulse  on the scale  $\sim 0.3 \mu \rm s$
could be   recorded with existing  technology as discussed in \cite{GUREVICH2003228}  in application to thunderstorm lightning events
where wide-band radio interferometry (0.1-30 MHz) has been used to detect such short signals with amplitudes of electric field on the level $10^2\rm ~mV/m$,   which is very close to  what is required for the purposes of the present work. Therefore, the predicted radio  signals    (\ref{E-numerics}) can be in principle   measured   if  proper instruments are designed and built at the TASD site.

Furthermore,  the short radio pulses can be, in principle,  recorded by any  (sufficiently sensitive) radio telescope outside of the TASD site as the corresponding radio pulses will be emitted in form of the clustering events, similar to TA-bursts. Such clusters of individual short radio pulses can be discriminated from   any spurious signals representing the  radio noise. It can be also discriminated from radio emission occurring as a result of  the thunderstorm lightning events as argued in Sect. \ref{mimic}.

Our  present proposal suggests  that the TA  bursts with very unusual features  will be synchronized  with radio pulses. 
If  this synchronization is observed and  the  interpretation of TA bursts as  the AQN annihilation events is confirmed by future studies
it would be the    {\it direct }  (non-gravitational) evidence which reveals  the nature of the DM, in contrast with large  number of {\it indirect}  hints mentioned in Sect. \ref{AQN}.

\begin{acknowledgements}
This research was supported in part by the Natural Sciences and Engineering
Research Council of Canada, and X.L. also by the University of British Columbia four year doctoral fellowship. 
\end{acknowledgements}

\bibliographystyle{utphys} 
\bibliography{Bursts-radio}

\end{document}